# Response of quantum pure states

**Barbara Fresch[1], Giorgio J. Moro[1]**

*Dipartimento di Science Chimiche, Università di Padova,*

*via Marzolo 1, 35131 Padova, Italy*

**Abstract**

The response of a quantum system in a pure state to an external force is investigated by reconsidering the standard statistical approach to quantum dynamics on the light of the statistical description of equilibrium based on typicality. We prove that the response of the large majority of quantum pure states subjected to the same arbitrary external perturbation tends to be close to the statistical response as the dimension of the Hilbert space increases. This is what we can term *dynamical typicality.* The theoretical analysis is substantiated by numerical simulations of the response of a spin system to a sudden quench of the external magnetic field. For the considered system we show that not only the system relaxes toward a new equilibrium state after the quench of the Hamiltonian but also that such a new equilibrium is compatible with the description of a *thermal* equilibrium.

[1] E-mails: barbara.fresch@unipd.it; giorgio.moro@unipd.it

# I. Introduction

One of the classic problems in non equilibrium statistical mechanics is to formulate the proper dynamic response function that applies when a system is subjected to a time dependent external field [1]. Traditionally, there are two methods to working out the response of a many body quantum system, one is the approach based on the formulation of a master equation for the density matrix [2,3] and the other relies on linear response theory [4]. The fundamental assumptions of both approaches are based on the traditional concept of statistical ensemble and ensemble averages which has been very controversial as physical principles. Recently however there has been significant progress in clarify foundational issue of quantum statistical mechanics relying on the evidence that *individual* quantum states of system can exhibit statistical properties. This approach has become a very fruitful direction of research in recent years [5] and it has established the concept of typicality as the key to the occurrence of standard statistical equilibrium behaviour (as opposed to ergodycity, mixing, etc.). In this paper we focus on the response to an external force of a single random pure state describing an isolated system. Starting from the exact quantum dynamical evolution of the system we shall develop a statistical theory of the response which is clearly related to the underlying mechanical description, without introducing ad hoc statistical assumptions. In such a framework it is easily demonstrated that a wide class of observables undergoes to a typical time evolution under the influence of an arbitrary external field. The plan of the paper is as follows: in Section II we recall the treatment of the equilibrium state for quantum pure states based on typicality and provide a detailed analysis of the temporal and statistical fluctuations of observables at the equilibrium. In Section III we demonstrate that a similar analysis can be apply to the response of a quantum pure state to an external perturbation. In particular we shall show that different random pure states characterized by the same thermodynamic properties yield very similar response in terms of a wide class of observables. In particular local observable, as those pertaining to subsystem properties, are shown to undergo a typical dynamics as a consequence of the application of an arbitrary external perturbation. In the second part of this section we also provide an illustration of the theory by numerical simulations of the response of a spin system to a sudden quench of the external magnetic field. For the considered system we show that not only the system relax toward a new equilibrium state after the quench of the Hamiltonian but also that such a new equilibrium is compatible with the description of a *thermal* equilibrium [6]. In Section IV the particular case of a linear response is considered and the corresponding form of the response function is derived and connected with the standard linear response by Kubo [4]. Finally, in the concluding section the main results of the present paper are summarized and commented.

## II. Statistical characterization of equilibrium properties

We shall first consider an isolated quantum system in a pure state characterized by a well defined wavefunction $|\psi(t)\rangle$ belonging to the associated Hilbert space $\mathcal{H}$. When it is let alone, such a system evolves in time according to the unitary Schrödinger evolution ruled by the system Hamiltonian $\hat{H}$. Notice that the condition of isolation is by far more stringent in quantum mechanics then in the classical case: indeed, even in the absence of any energetic interactions, a quantum system can always undergo entangling interactions with the surrounding which prevent the possibility to assign it a well defined wavefunction [7,8,9].

By solving the equation of motion one can, at least in principle, determine the complete dynamics of the system. Then the time evolution can be analyzed in terms of the probability distribution on a set of parameters specifying the wavefunction at any instant. Such an approach has been extensively discussed in Refs. [10,11] to obtain the statistical characterization of the equilibrium properties of composite quantum systems. In the following we shall recall the definition and the main results which are invoked in the subsequent analysis of the response to an external force. Moreover, we shall present some new results about typicality and the amplitude of quantum fluctuations.

*Quantum Dynamics and the Pure State Distribution*. The central point of equilibrium statistical mechanics is the replacement of the mechanical description given by the time evolution of the state of the system, with a description in terms of probability density on the space which represents the possible states of the system during its motion [12]. According to quantum mechanics the time evolution of the isolated systems characterized by the time independent Hamiltonian $\hat{H}$, is described by the Schrödinger equation

$$\frac{\partial}{\partial t}|\psi(t)\rangle = -\frac{i}{\hbar}\hat{H}|\psi(t)\rangle \qquad (1)$$

having an unique solution $|\psi(t)\rangle = \exp(-i\hat{H}t/\hbar)|\psi(0)\rangle$ once the state $|\psi(0)\rangle$ at $t=0$ is specified. An equivalent description is obtained by means of the pure state density operator

$$\hat{\rho}(t) := |\psi(t)\rangle\langle\psi(t)| \qquad (2)$$

leading to a linear relationship for the time dependent expectation value $B(t) := \langle\psi(t)|\hat{B}|\psi(t)\rangle$ of any property associated to a (self-adjoint) operator $\hat{B}$

$$B(t) = \mathrm{Tr}\{\hat{B}\hat{\rho}(t)\} \quad . \qquad (3)$$

By choosing a particular subsystem $S$, and the environment $E$ as the rest of the overall system, one can factorize the Hilbert space with respect to the corresponding subspaces $\mathcal{H} = \mathcal{H}_S \otimes \mathcal{H}_E$. If an operator $\hat{b}$ of the subsystem $S$ is considered, then only the reduced density matrix $\hat{\mu}(t)$ traced out on the environment states

$$\hat{\mu}(t) := \text{Tr}_E \{\hat{\rho}(t)\} \tag{4}$$

is required to evaluate the corresponding expectation value $b(t)$

$$b(t) = Tr\{(\hat{b} \otimes \hat{1}_E)\hat{\rho}(t)\} = Tr_S\{\hat{b}\hat{\mu}(t)\} \tag{5}$$

Calculations for large enough systems with Hamiltonian having at least partially a random character display for the observables (expectation values) a fluctuating behaviour about the average, typical of stochastic variables (see, for example, Fig. 1 of ref. [11]). This suggests that the average and the amplitude of fluctuations about the average have to be considered as the two main parameters characterizing the statistical properties developed in the time by a given observable $B(t)$. The equilibrium average is defined through the asymptotic time average

$$\overline{B} := \lim_{T \to \infty} \frac{1}{T} \int_0^T dt\, B(t) = \text{Tr}(\hat{B}\overline{\hat{\rho}}) \tag{6}$$

which is completely determined by the average density matrix

$$\overline{\hat{\rho}} := \lim_{T \to \infty} \frac{1}{T} \int_0^T dt\, \hat{\rho}(t) \tag{7}$$

In the case of isolated systems evolving under their own Hamiltonian, the over-bar on a given property will be used to denote the equilibrium average deriving from the asymptotic time average of the corresponding pure state evolution. The natural parameter quantifying the fluctuation amplitude of an observable $B(t)$ is the mean squared deviation from the average

$$\overline{\Delta B^2} := \overline{(B(t) - \overline{B})^2} = \lim_{T \to \infty} \frac{1}{T} \int_0^T dt\, \left[ \text{Tr}\left(\hat{B}(\hat{\rho}(t) - \overline{\hat{\rho}})\right) \right]^2 \tag{8}$$

A deeper analysis of the statistics of quantum pure states requires the identification of a convenient set of parameters which univocally specifies the state of the system at any time, and which can be used as coordinates for the phase space and, therefore, as independent variables for the probability density induced by the evolution of the quantum state. Once such a probability density is available, the equilibrium averages, for instance those of eq. (6) and of eq. (8), can be determined directly by integration on the phase space. This problem has been addressed in refs. [10,11] and we report here only the general set up and the results.

Let us introduce the eigenbasis of the system Hamiltonian $\hat{H}|E_k\rangle = E_k|E_k\rangle$, with $k = 1, 2, \ldots$. In order to deal with a finite set parameterization of the wavefunction, we assume that $|\psi(t)\rangle$ belongs to the finite dimensional subspace $\mathcal{H}_N \subseteq \mathcal{H}$, in the following called as the active Hilbert space for the wavefunction, defined on the basis of the cut-off energy $E_{\max}$

$$\mathcal{H}_N := \mathrm{span}\{|E_k\rangle | E_k < E_{\max}\} \tag{9}$$

where $N$ is its dimension: $E_N < E_{\max} \leq E_{N+1}$. In the energy eigen-basis the time evolution of the density operator eq. (2) can be written explicitly as

$$\hat{\rho}(t) = \sum_{k=1}^{N} \sum_{k'=1}^{N} c_k(0) c_{k'}(0)^* \exp[-i(E_k - E_{k'})t/\hbar] |E_k\rangle\langle E_{k'}| \tag{10}$$

where $c_k(t) := \langle E_k | \psi(t) \rangle$ are the (complex) components of the wavefunction in the energy representation. Notice that the diagonal elements of the pure state density matrix eq. (10) do not depend on time, while the off-diagonal terms are oscillating functions of the time in the absence of degeneracy in the energy spectrum. As discussed in ref. [10,11], the random effects of intermolecular interactions justifies the assumption of rational independence [13] of energy eigenvalues (also called the non-resonance condition). Such a condition, implying the absence of degeneracy in the energy spectrum of the system, will be assumed in our analysis. Accordingly the equilibrium average density matrix eq. (7) is explicitly given as

$$\overline{\rho} = \sum_{k=1}^{N} P_k |E_k\rangle\langle E_k| \tag{11}$$

where we have introduced the populations of energy eigenstates $P_k = |c_k(0)|^2$ for $n = 1, 2, \ldots N$, normalized as $\sum_{k=1}^{N} P_k = 1$. They represent the constants of motion for the quantum dynamics. The rational independence of the energy eigenvalues allows the direct evaluation of the time averages in eq. (6) and eq. (8), leading to an explicit dependence on the populations for the averages and the fluctuation amplitudes

$$\overline{B} = \sum_{k=1}^{N} B_{k,k} P_k \tag{12}$$

$$\overline{\Delta B^2} = \sum_{k,k'=1}^{N} B_{k,k'} B_{k',k} P_k P_{k'} - \sum_{k=1}^{N} B_{k,k}^2 P_k^2 \tag{13}$$

with $B_{k,k'} := \langle E_k | \hat{B} | E_{k'} \rangle$.

By introducing the phases $\alpha_k(t)$ for the polar representation of the wavefunction coefficients, $c_k(t) = |c_k(t)| \exp[i\alpha_k(t)]$, which depend linearly on the time

$$\alpha_k(t) = \alpha_k(0) - E_k t/\hbar \tag{14}$$

we can parameterize the pure state density operator by means of the constant populations and the time dependent phases

$$\hat{\rho}(t) = \sum_{k,k'=1}^{N} \sqrt{P_k P_{k'}} \exp(i\alpha_k(t) + i\alpha_{k'}(t)) |E_k\rangle\langle E_{k'}| \tag{15}$$

Correspondingly also the instantaneous value of the observable $B(t)$ eq. (3) can be parameterized according to the same set of variables. Then, for a given set of populations $P = (P_1, P_2, \cdots, P_N)$, we can consider the phases $\alpha = (\alpha_1, \alpha_2, \cdots, \alpha_N)$ as the statistical variables which determine the instantaneous state of the system. By introducing the probability density $p_P(\alpha)$ on the phases, in principle parametrically dependent on the populations, one can evaluate the phase space average of any observable to be considered as a phase function $f_P(\alpha)$, parametrically dependent on the populations, by integration on the phases $\int_0^{2\pi} d\alpha_1 \int_0^{2\pi} d\alpha_2 \cdots \int_0^{2\pi} d\alpha_N f_P(\alpha) p_P(\alpha)$. By imposing the condition of equivalence between time average and phase average one can determine the probability density $p_P(\alpha)$. In Refs [10] it has been shown that for a rational independent energy spectrum [13], such a probability density, called by as the Pure State Distribution (PSD), is given as a population independent and homogeneous function of the phases [14]:

$$p_P(\alpha) = \frac{1}{(2\pi)^N} \tag{16}$$

In conclusion, the asymptotic time average has to be considered equivalent to a simple average on the phases when equilibrium properties of an isolated quantum system are considered.

*Population statistics with the Random Pure State Ensemble.* By defining the equilibrium average of an observable on the basis of the dynamics of the pure state, as done in the previous section, one does not need any a priori assumption, as e.g. *ad hoc* definition of the microcanonical or the canonical density matrix, to represent the equilibrium state of the system. On the other hand any equilibrium average, like in eq.(12) and eq. (13), depends on the specific set of populations of the pure state describing a specific realization of the considered system.

As a matter of fact the empirical, full knowledge of the population set for a system with several degrees of freedom is impossible. One has to resort to a statistical analysis on the basis of an ensemble [15] specified by the sample space and the probability density $p(P)$ for the populations. In the choice of the statistical ensemble, an essential requirement is the agreement with the thermodynamic description which specifies the equilibrium state of a macroscopic system on the basis of few macroscopic variables [10,11]. To this aim, one has to recognize that the details of the population set $P$ are largely irrelevant in determining the macroscopic equilibrium properties as long as the pure state belongs to a high dimensional Hilbert space [16]. In the recent literature [5,6,16], typicality is attributed to a property whenever it has nearly the same value in the overwhelming majority of the pure states allowed to the system. The population density $p(P)$ allows a rigorous definition of typicality. Let us consider an equilibrium property $g(P)$ which, like in eqs. (12) and (13), depends on the population set. Then, once the statistical ensemble has been specified, one can calculate its ensemble average (in the following denoted as the function between the bracket)

$$\langle g \rangle := \int dP\, p(P) g(P) \tag{17}$$

Analogously one can define the variance of the property $g(P)$ within the ensemble

$$\sigma_g := \sqrt{\langle (g - \langle g \rangle)^2 \rangle} \tag{18}$$

Then the statement that the overwhelming majority of pure states has nearly the same property, can be quantified by the requirement that the range of its statistically significant values, which is determined according to $\sigma_g$, is much smaller than the domain size $\Delta_g$ of all its possible values: $\sigma_g \ll \Delta_g$. In establishing the connection with thermodynamics, a stronger (asymptotic) form of typicality is invoked by requiring that the ratio $\sigma_g / \Delta_g$ tends to vanish in the limit of an infinite dimension of the active space, in correspondence of the macroscopic limit for the considered system

$$\lim_{N \to \infty} \sigma_g / \Delta_g = 0 \tag{19}$$

In ref. [17] we have introduced the Random Pure State Ensemble (RPSE) corresponding to a random choice of the wavefunction within the unitary hyper-sphere in the active Hilbert space, eq. (9). The resulting probability density on the populations is an uniform distribution on the $(N-1)$ dimensional simplex determined by the normalization condition $\sum_{k=1}^{N} P_k = 1$, leading to the following relations for the average and the variance (within the ensemble) of a population

$$\langle P_k \rangle = \frac{1}{N} \qquad \sigma_{P_k}^2 := \langle (P_k - \langle P_k \rangle)^2 \rangle = \frac{N-1}{(N+1)N^2} \tag{20}$$

Furthermore, the correlation coefficient between two populations is given as

$$k \neq k': \quad \frac{\langle (P_k - \langle P_k \rangle)(P_{k'} - \langle P_{k'} \rangle) \rangle}{\sigma_{P_k} \sigma_{P_{k'}}} = -\frac{1}{(N-1)} \tag{21}$$

The key parameter of the RPSE is the energy cut-off $E_{\max}$, which determines the dimension $N = N(E_{\max})$ of the RPSE active space. For the analysis of the typicality of expectation values presented in the following, it is important to determine how such a parameter scales with the size of the considered system. To this aim let us specifically consider a "modular" system, that is, a system which can be divided into $n$ subunits or components. For such a kind of systems we have shown in ref. [11] that, in the limit of large $n$, both the energy, $E := \langle \psi | H | \psi \rangle = \sum_{k=1}^{N} E_k P_k$, and the Shannon entropy with respect to the populations, $S := -k_B \sum_{k=1}^{N} P_k \ln P_k$, satisfy the condition eq. (19) of typicality, and that their average within the Ensemble can be identified with the thermodynamic internal energy $U$ and entropy $S$, respectively. It turns out that the dimension of the active Hilbert space is directly related to the thermodynamic state of the isolated system as

$$N = \exp(ns(u)) \tag{22}$$

where $s(u)$ is the entropy per component $s := S/n$ as a function of the energy per component $u := U/n$. In conclusion, for a given temperature determining the energy per component $u$, and therefore also $s$, the dimension $N$ of the active Hilbert space grows exponentially with the size $n$ of the isolated system. It should be recalled [11] that in the same limit the typical equilibrium reduced density matrix $\langle \overline{\hat{\mu}} \rangle$ assumes the canonical form

$$\langle \overline{\hat{\mu}} \rangle = \frac{\exp(-\hat{H}_S / k_B T)}{\text{Tr}_S \{\exp(-\hat{H}_S / k_B T)\}} \tag{23}$$

where $\hat{H}_S$ is the subsystem Hamiltonian, provided that the interaction Hamiltonian between subsystem and environment has negligible effects.

*Amplitude of fluctuations and typicality of equilibrium properties.* In this subsection, we shall provide explicit estimates for the fluctuation amplitude in the PSD and the RPSE variance for the observables. They are straightforward consequences of the analysis by Gemmer et al. [18,20] of the variance of expectation values for randomly distributed pure quantum states in the Hilbert space. However, we think that a deeper insight is gained within our framework since the consequences of the randomness of the wavefunction in the Hilbert space is substantiated into separate effects of time fluctuations and of different realizations of the pure quantum state. As applications, subsystem properties will be considered.

Let us first perform the typicality analysis of the generic property $\overline{B}$ of eq. (6). By taking into account eq. (20), its RPSE average is given as

$$\langle \overline{B} \rangle = \frac{1}{N} \sum_{k=1}^{N} B_{k,k} = \frac{1}{N} \text{Tr}\{\hat{B}_N\} \tag{24}$$

where

$$\hat{B}_N := \hat{Q}_N \hat{B} \hat{Q}_N \tag{25}$$

is operator $\hat{B}$ projected onto $\mathcal{H}_N$ according to the projection operator

$$\hat{Q}_N := \sum_{k=1}^{N} |E_k\rangle\langle E_k| \qquad (26)$$

Also the variance within RPSE of such an equilibrium property

$$\sigma_{\overline{B}}^2 = \langle (\overline{B} - \langle \overline{B} \rangle)^2 \rangle = \sum_{k=1}^{N}\sum_{k'=1}^{N} B_{k,k} B_{k',k'} \langle (P_k - \langle P_k \rangle)(P_{k'} - \langle P_{k'} \rangle) \rangle \qquad (27)$$

can be evaluated according to population statistics eqs. (20) and (21), so obtaining

$$\sigma_{\overline{B}}^2 = \frac{1}{N+1}\left( \frac{\sum_{k=1}^{N} B_{k,k}^2}{N} - \frac{\text{Tr}\{\hat{B}_N\}^2}{N^2} \right) \qquad (28)$$

On the other hand the RPSE average of the fluctuation amplitude eq. (13) can be specified as

$$\langle \overline{\Delta B^2} \rangle = \frac{1}{N(N+1)}\left( \sum_{k,k'=1}^{N} B_{k,k'} B_{k',k} - \sum_{k=1}^{N} B_{k,k}^2 \right) = \frac{1}{N(N+1)}\left( \text{Tr}\{\hat{B}_N^2\} - \sum_{k=1}^{N} B_{k,k}^2 \right) \qquad (29)$$

Therefore, within the RPSE, the variance and the fluctuation amplitude of a generic equilibrium property $\overline{B}$ can be summed up as

$$\sigma_{\overline{B}}^2 + \langle \overline{\Delta B^2} \rangle = \frac{1}{N+1}\left( \frac{\text{Tr}\{\hat{B}_N^2\}}{N} - \frac{\text{Tr}\{\hat{B}_N\}^2}{N^2} \right) \qquad (30)$$

As a matter of fact the right end side is determined by the spectral distribution of $\hat{B}_N$. Let us denote the eigenvalues of $\hat{B}_N$ as $\lambda_1, \lambda_2, \cdots, \lambda_N$. Then the average spectral value of $\hat{B}_N$ can be specified as

$$D_1(\hat{B}_N) := \sum_{k=1}^{N} \lambda_k / N = \text{Tr}\{\hat{B}_N\}/N \qquad (31)$$

while the squared spectral variance of $\hat{B}_N$ is given as

$$D_2(\hat{B}_N) := \sum_{k=1}^{N} (\lambda_k - D_1(\hat{B}_N))^2 / N = \sum_{k=1}^{N} \frac{\lambda_k^2}{N} - D_1(\hat{B}_N)^2 = \frac{\text{Tr}\{\hat{B}_N^2\}}{N} - \frac{\text{Tr}\{\hat{B}_N\}^2}{N^2} \qquad (32)$$

In conclusion eq. (30) finds the following more transparent form

$$\sigma_{\overline{B}}^2 + \left\langle \overline{\Delta B^2} \right\rangle = \frac{D_2(\hat{B}_N)}{N+1} \tag{33}$$

Notice that by considering together the RPSE average $\left\langle \overline{\Delta B^2} \right\rangle$ of fluctuation amplitude and the RPSE squared variance $\sigma_{\overline{B}}^2$ of the equilibrium property $\overline{B}$, one recovers the cumulative mean squared deviation of the observable $B(t)$ from its RPSE average equilibrium value $\left\langle \overline{B} \right\rangle$

$$\left\langle \overline{[B(t) - \langle \overline{B} \rangle]^2} \right\rangle = \left\langle \overline{[\Delta B(t) - (\overline{B} - \langle \overline{B} \rangle)]^2} \right\rangle = \left\langle \overline{\Delta B^2} \right\rangle + \left\langle (\overline{B} - \langle \overline{B} \rangle)^2 \right\rangle = \left\langle \overline{\Delta B^2} \right\rangle + \sigma_{\overline{B}}^2 \tag{34}$$

This corresponds to the mean squared deviation of the observable from the average, by taking into account both the distribution on the phases and on the populations, that is the result for the statistics of a random distribution of the wavefunction within the unit sphere in the Hilbert space, or in one of its subspaces [17]. From this point of view, eq. (33) is equivalent to the result of Gemmer et al. [18,20] that $D_2(\hat{B})/(N+1)$ is the Hilbert space squared variance for the expectation value of operator $\hat{B}$. However, the form of eq. (33) separates the effects of fluctuations in time measured by $\overline{\Delta B^2} = \overline{(B - \overline{B})^2}$ in a single realization of the system, from the squared variance $\sigma_{\overline{B}}^2 = \left\langle (\overline{B} - \langle \overline{B} \rangle)^2 \right\rangle$ which could detected only by comparing the equilibrium average $\overline{B}$ in different realizations of the system.

Equation (33) has some important consequences since, on the basis of simple spectral spectral properties of operator $\hat{B}_N$, it determines an upper bound to both the average fluctuation amplitude $\left\langle \overline{\Delta B^2} \right\rangle$ and the variance $\sigma_{\overline{B}}$ of equilibrium property $\overline{B}$. Let us consider the asymptotic limit with respect to the size of the system. By considering the system as an ensemble of identical components or subunits at a given temperature, according to eq. (22) this corresponds to the limit $N \to \infty$. Thus if operator $\hat{B}$ has a bounded spectrum, that is if two constants $B_{min}$ and $B_{max}$ exist such that the following condition holds

$$B_{min} \leq \frac{\langle \psi | \hat{B} | \psi \rangle}{\langle \psi | \psi \rangle} \leq B_{max} \tag{35}$$

for any wavefunction $|\psi\rangle$, the spectral variance of $\hat{B}_N$ has the following upper bound independent of the dimension $N$ of the active space

$$D_2(\hat{B}_N) < (B_{\max} - B_{\min})^2 \qquad (36)$$

and, therefore, a vanishing value for the asymptotic values of both the fluctuation amplitude and the variance is recovered from eq. (33)

$$\lim_{N\to\infty}\overline{\langle \Delta B^2\rangle} = \lim_{N\to\infty} \sigma_{\overline{B}} = 0 \qquad (37)$$

Therefore, in the thermodynamic limit, not only the typicality eq. (19) for the equilibrium value $\overline{B}$ of the observable $B(t)$, but also vanishing amplitude for its fluctuations is predicted.

An obvious application is that for spin observable having an intrinsically bounded spectrum. By properly identifying operator $\hat{B}$, the same method can be applied to demonstrate the typicality and the vanishing of fluctuations of the reduced density matrix $\hat{\mu}(t)$ eq. (4) in the thermodynamic limit. Let us denote by $|m\rangle$ for $m = 1,2,\cdots$ the elements of an arbitrary orthonormal basis of the Hilbert subspace $\mathcal{H}_S$ for the subsystem. The $m$-th diagonal element of $\hat{\mu}(t)$ can be identified with the expectation value $b(t) = \mu_{m,m}(t) := \langle m|\hat{\mu}(t)|m\rangle$ of subsystem operator $\hat{b} = |m\rangle\langle m|$, i.e. operator $\hat{B} = \hat{b}\otimes\hat{1}_E$ in the full Hilbert space, which has $0$ and $1$ has possible eigenvalues. Since $\hat{B}$ has a bounded spectrum, then one can apply the limits eq. (37) to the diagonal elements of the reduced density matrix. In a similar way one can analyze the off-diagonal elements by choosing $\hat{b} = (|m'\rangle\langle m| + |m\rangle\langle m'|)/2$ and $\hat{b} = (|m'\rangle\langle m| - |m\rangle\langle m'|)/2i$ as the subsystem operator whose expectation value is the real part and the imaginary part, respectively, of $\mu_{m,m'}(t)$. Also in this case operator $\hat{B} = \hat{b}\otimes\hat{1}_E$ is bounded, $0$ and $\pm 1/2$ being its eigenvalues, and the asymptotic condition eq. (37) is again satisfied. Notice that also in ref. [11] we have verified the typicality of the reduced density matrix in the thermodynamic limit, but by assuming a negligible Hamiltonian of interaction between the subsystem and the environment. Evidently the present analysis is more general since no condition on the interaction Hamiltonian is required.

Finally we evaluate the RPSE average of the fluctuation amplitude $\overline{\langle \Delta b^2\rangle}$ and the RPSE variance $\sigma_{\overline{b}}$ for the expectation value $b(t) = \operatorname{Tr}_S\{\hat{b}\hat{\mu}(t)\}$ for a generic (self-adjoint) operator $\hat{b}$ of the subsystem, with equilibrium value $\overline{b} = \mathrm{Tr}_S\{\hat{b}\overline{\hat{\mu}}\}$. Notice that the RPSE average $\langle\overline{\hat{\mu}}\rangle = \mathrm{Tr}_E\{\langle\overline{\hat{\rho}}\rangle\}$ of the equilibrium reduced density matrix can be specified according to projection operator eq. (26)

$$\langle\overline{\hat{\mu}}\rangle = \frac{1}{N}\mathrm{Tr}_E\{\hat{Q}_N\} \tag{38}$$

Equation (30), with $\overline{B} = \overline{b}$ and $\Delta B = \Delta b = b(t) - \overline{b}$ can be applied also the present case once operator $\hat{B}$ is identified with $\hat{b}\otimes\hat{1}_E$. Furthermore the trace on $\hat{B}_N$ can be specified as

$$\frac{\mathrm{Tr}\{\hat{B}_N\}}{N} = \frac{\mathrm{Tr}\{(\hat{b}\otimes\hat{1}_E)\hat{Q}_N\}}{N} = \mathrm{Tr}_S\{\hat{b}\mathrm{Tr}_E\{\hat{Q}_N/N\}\} = \mathrm{Tr}_S\{\hat{b}\langle\overline{\hat{\mu}}\rangle\} \tag{39}$$

By taking into account that, for any operator $\hat{B}$, $\mathrm{Tr}\{\hat{Q}_N\hat{B}^2\hat{Q}_N\} - \mathrm{Tr}\{\hat{Q}_N\hat{B}\hat{Q}_N\hat{B}\hat{Q}_N\} \geq 0$, as one can easily verify by substitution of projection operator eq. (26), the trace of squared $\hat{B}_N$ has the following upper bound

$$\frac{\mathrm{Tr}\{\hat{B}_N^2\}}{N} \leq \frac{\mathrm{Tr}\{\hat{Q}_N\hat{B}^2\hat{Q}_N\}}{N} = \frac{\mathrm{Tr}\{(\hat{b}^2\otimes 1_E)\hat{Q}_N\}}{N} = \mathrm{Tr}_S\{\hat{b}^2\mathrm{Tr}_E\{\hat{Q}_N/N\}\} = \mathrm{Tr}_S\{\hat{b}^2\langle\overline{\hat{\mu}}\rangle\} \tag{40}$$

In conclusion eq. (33) can be substituted by the inequality

$$\langle\overline{\Delta b^2}\rangle + \sigma_{\overline{b}}^2 \leq \frac{\mathrm{Tr}_S\{\hat{b}^2\langle\overline{\hat{\mu}}\rangle\} - \mathrm{Tr}_S\{\hat{b}\langle\overline{\hat{\mu}}\rangle\}}{N+1} \tag{41}$$

with the right hand side representing an upper bound to both average squared amplitude of fluctuations and the squared RPSE variance. For large enough systems the RPSE average $\langle\overline{\hat{\mu}}\rangle$ of the equilibrium reduced density matrix becomes independent of the dimension $N$ of the Hilbert active space, like the canonical form eq. (23). Then the upper bound in eq. (41) results to be inversely proportional to $(N+1)$ and, in the asymptotic limit for increasing system size, both the fluctuation amplitude and the RPSE variance of equilibrium subsystem properties vanish

$$\lim_{N\to\infty}\langle\overline{\Delta b^2}\rangle = \lim_{N\to\infty}\sigma_{\overline{b}} = 0 \tag{42}$$

In other words, if the dimension of the active space is sufficiently large, the instantaneous value of any subsystem property is practically equal to its equilibrium value $b(t) \cong \overline{b}$ for most of the time. Moreover, because of typicality, $\sigma_{\overline{b}} \to 0$, the equilibrium value is nearly coincident with the canonical average $\langle \overline{b} \rangle = \text{Tr}_S \{\hat{b} \langle \overline{\hat{\mu}} \rangle\}$ calculated according to eq. (23). One way for observing time dependent phenomena requires the action from the outside with an external force. This situation will be considered in the next section.

### III. Response to an external force

In the previous section we have outlined the statistical description of a pure state which evolves according to the Schrödinger equation ruled by the system Hamiltonian. We now turn to our main topic, the response of a pure state to an external force. In the first part of this section we shall treat in all generality the response of the system. In particular it will be shown that the typicality properties previously derived for of the isolated system can be generalized to the response to an external force. In the second part of this section, some results for the quench of the magnetic field acting on a spin system will be presented in order to illustrate an application of the theory.

**III-A General Theory**

Let $F(\tau)$ for $\tau \geq 0$ be the time profile of an external field which is turned on at a time $t_F$. The Hamiltonian which rules the dynamics is written as

$$\hat{H}(t) = \hat{H} + \hat{A} F(t - t_F) \qquad (43)$$

where $H$ is the Hamiltonian of the isolated system, while $\hat{A}$ is the system operator for the coupling with the external field. From the Schrödinger equation including the external force

$$i\hbar \frac{\partial}{\partial t} |\psi(t)\rangle = \hat{H} |\psi(t)\rangle + F(t - t_F) \hat{A} |\psi(t)\rangle \qquad (44)$$

one can derive the unitary operator $\hat{U}(\tau)$ determining the wavefunction after a time $\tau$ from the application of the force

$$|\psi(t_F + \tau)\rangle = U(\tau) |\psi(t_F)\rangle \qquad (45)$$

with $\hat{U}(0) = \hat{1}$. An explicit form for the evolution operator $\hat{U}(\tau)$ can be provided as a time ordered exponential [19]. The previous description of the pure state (PSD) and its statistics (RPSE) will be assumed for the system before the application of the external force, with the wavefunction $|\psi(t)\rangle$ for $t \leq t_F$ belonging to the active Hilbert space $\mathcal{H}_N$. Notice that in all generality operator $\hat{A}$ in the time dependent Hamiltonian eq. (43) introduces a coupling between $\mathcal{H}_N$ and its complementary space in $\mathcal{H}$ and, therefore, for $t > t_F$ the wavefunction $|\psi(t)\rangle$ would have components outside $\mathcal{H}_N$.

Let us now consider as observable the expectation value $B(t_F, \tau)$ of an operator at time $\tau$ after the application at time $t_F$ of the external force

$$B(t_F,\tau) := \langle \psi(t_F+\tau)|\hat{B}|\psi(t_F+\tau)\rangle = \langle \psi(t_F)|\hat{B}(\tau)|\psi(t_F)\rangle = \text{Tr}\{\hat{B}(\tau)\hat{\rho}(t_F)\} \qquad (46)$$

where $\hat{\rho}(t_F) = |\psi(t_F)\rangle\langle\psi(t_F)|$ is the density matrix just before the application of the force, and

$$\hat{B}(\tau) := \hat{U}(\tau)^\dagger \hat{B}\hat{U}(\tau) \qquad (47)$$

is the operator in the Heisenberg representation. The observable $B(t_F, \tau)$ can be used to characterize the response of the system to the external force. Specific results reported further on for specific cases, show that such a response can be interpreted as relaxation induced by a change in the external field but also display fluctuations typical of the evolution of isolated systems. In order to clearly recognize the relaxation behaviour in the response $B(t_F, \tau)$, the fluctuations have to be eliminated, like in the calculation of equilibrium averages for the pure state of an isolated system. This could be done by averaging the response over the phases of the wavefunction $|\psi(t_F)\rangle$ at the time of application of the force. However, as shown in the analysis of the Pure State Distribution (PSD), this is equivalent to perform the average on the time dependence of $|\psi(t_F)\rangle$. Therefore we introduce the following asymptotic average on the time of the force application, in order to define the response $\overline{B(\tau)}$ averaged over the fluctuations

$$\overline{B(\tau)} := \lim_{T\to\infty} \frac{1}{T} \int_0^T dt_F\, B(t_F,\tau) = \text{Tr}\left\{\hat{B}(\tau)\lim_{T\to\infty}\frac{1}{T}\int_0^T dt_F\, \hat{\rho}(t_F)\right\} = \text{Tr}\{\hat{B}(\tau)\overline{\hat{\rho}}\} \qquad (48)$$

where $\overline{\hat{\rho}}$ is the equilibrium (averaged) density matrix of the isolated system before the application of the external force. The amplitude of fluctuations at a given time $\tau$ after the application of the force can be quantified by evaluating the mean squared deviation like for isolated systems

$$\overline{\Delta B(\tau)^2} := \overline{\left[B(t_F,\tau)-\overline{B(\tau)}\right]^2} = \lim_{T\to\infty}\frac{1}{T}\int_0^T dt_F \left(\text{Tr}\{\hat{B}(\tau)[\hat{\rho}(t_F)-\overline{\hat{\rho}}]\}\right)^2 \qquad (49)$$

Even if with the previous procedure the dependence on the phases of the wavefunction $|\psi(t_F)\rangle$ at the force application is eliminated, still the response depends on the particular realization of the populations. One can use RPSE statistics to study such a population dependence of the fluctuation's mean response $\overline{B(\tau)}$, and to define the average over the possible realizations of the pure state

$$\langle \overline{B(\tau)} \rangle = \text{Tr}\{\hat{B}(\tau)\langle\overline{\hat{\rho}}\rangle\} \qquad (50)$$

where the bracket $\langle\cdots\rangle$ will continue to denote the RPSE average on the populations like in the previous section. Since $\langle\overline{\hat{\rho}}\rangle$ is the RPSE averaged equilibrium density matrix prior the force application, we can employ the results of the previous section for the isolated system, and in particular eq. (24) to derive the relation

$$\left\langle \overline{B(\tau)} \right\rangle = \frac{1}{N} \text{Tr}\left\{ \hat{B}_N(\tau) \right\} \qquad (51)$$

where $\hat{B}_N(\tau)$ is operator $\hat{B}(\tau)$ projected onto the active Hilbert space $\mathcal{H}_N$

$$\hat{B}_N(\tau) := Q_N \hat{B}(\tau) Q_N \qquad (52)$$

In the following we shall present some numerical calculations of the magnetization dynamics of a spin system subjected to a quantum quench of the static magnetic field. In figure 5A the fluctuation averaged response of five different pure states drawn from the same RPSE are reported (coloured lines) together with its RPSE average (black line). It is evident that the statistical response is quite insensible to the different choice of the pure state populations. To quantify such dependence we should compute the variance of the statistical response within the RPSE statistics

$$\sigma_{\overline{B(\tau)}}^2 = \left\langle \left( \overline{B(\tau)} - \left\langle \overline{B(\tau)} \right\rangle \right)^2 \right\rangle = \frac{1}{N+1} \left( \frac{\sum_{k=1}^N \hat{B}_{k,k}(\tau)^2}{N} - \frac{\text{Tr}\{\hat{B}(\tau)\}^2}{N^2} \right) \qquad (53)$$

Like for eq. (29), one can derive the following relation for the RPSE average of fluctuation amplitude

$$\left\langle \overline{\Delta B(\tau)^2} \right\rangle = \frac{1}{N(N+1)} \left( \text{Tr}\{\hat{B}_N(\tau)^2\} - \sum_{k=1}^N B_{k,k}(\tau)^2 \right) \qquad (54)$$

In practice all the statistical analysis about typicality of pure states of isolated systems, which has been reported in the previous section, can be transferred to the response to an external force by substituting operator $\hat{B}$ with its Heisenberg form $\hat{B}(\tau)$ eq. (47). In particular one recovers the counterpart of eq. (33) as the constraint for the RPSE variance and the amplitude of fluctuations

$$\sigma_{\overline{B(\tau)}}^2 + \left\langle \overline{\Delta B(\tau)^2} \right\rangle = \frac{D_2(\hat{B}_N(\tau))}{N+1} \qquad (55)$$

Then, if operator $\hat{B}$ is bounded according to eq. (35), also its Heisenberg form is bounded since $\langle \psi | \hat{B}(\tau) | \psi \rangle / \langle \psi | \psi \rangle = \langle \psi' | \hat{B} | \psi' \rangle / \langle \psi' | \psi' \rangle$ with $|\psi'\rangle = \hat{U}(\tau)|\psi\rangle$, and in the large size limit of the system both the typicality of the time dependent response and the vanishing of fluctuations is assured

$$\lim_{N \to \infty} \left\langle \overline{\Delta B(\tau)^2} \right\rangle = \lim_{N \to \infty} \sigma_{\overline{B(\tau)}} = 0 \qquad (56)$$

From such a result, by employing the same procedures described in the previous section for the isolated system and by recognizing the properties shared by $\hat{B}$ and $\hat{B}(\tau)$, one can demonstrate in the thermodynamic limit the typicality and the vanishing of fluctuation amplitude for the reduced density matrix and the expectation values of a subsystem at any time $\tau$.

In conclusion typicality is not only a feature of large isolated quantum systems, but also of their response to an external force. Correspondingly typicality acquires a dynamical perspective since by means of the response one can detect the relaxation behaviour with respect to external perturbation with a well defined time dependence. This is another manifestation of dynamical typicality, originally referred by Bartsch and Gemmer [20] to the evolution of isolated quantum systems from particular non-equilibrium initial conditions.

**III-B Response to a Quench of the Magnetic Field in a spin model system.**

The simplest setting to study the response of many body systems is to consider an abrupt change in time of one of the control parameters, i.e., a quantum quench. Thus we study the dynamics of a system composed of $n=10$ spins $J=1/2$ (two level system) when the external magnetic field suddenly changes from an initial value $B_i$ to a five times smaller value $B_f = B_i/5$. The total Hilbert space $\mathcal{H}$ has dimension $d=2^n$. For simplicity we assume the spins are all identical and the external field does not present any inhomogeneities so that the non interacting part of the Hamiltonian is given by $H_0 = H_S + H_E = \omega_0 \sum_{k=1}^{n} S_z^{(k)}$, where the Zeeman frequency $\omega_0$ is determined by the initial field. The eigenstates of this Hamiltonian will be denoted as $|se\rangle$ and we shall identify one spin as the (sub) system of interest S while the $n-1$ remaining spins act as the environment. Since we are not interest in the dynamical effects of some specific interactions but rather to the presence of a generic interaction between the components we assume that the product state basis is in no way a preferred basis for the interaction $H_{SE}$. Thus, we model the interaction Hamiltonian as a $d \times d$ Gaussian Orthogonal Random Matrix $W$ in the $|se\rangle$ representation. Such a matrix is a realization of the Gaussian Orthogonal Ensemble [21] whose probability density $P_d(W) = C\exp\left(-Tr\{W^2\}/2\sigma_W^2\right)$ is fully characterized by the dimension $d$ and the parameter $\sigma_W$. The $d(d+1)/2$ independent elements of a GORM are Gaussian random numbers with the following statistical properties

$$\langle W_{ij} \rangle_{GOE} = 0 \qquad \langle W_{ij}^2 \rangle_{GOE} = \frac{\sigma_W^2}{2}\left(1+\delta_{ij}\right) \tag{57}$$

we set $\sigma_W = 1$ while the interaction Hamiltonian is defined as

$$H_{SE} = \lambda W \tag{58}$$

where $\lambda$ is a control parameter assuring that $H_{SE}$ acts like a small perturbation to $H_0$, that is $|H_{SE}| \ll |H_0|$.

In the following we shall use such a simple model to illustrate the salient points of our treatment of the response. The natural observable to consider is the magnetization along the static magnetic field, thus we shall calculate the time evolving single spin polarization, $m_z(t) = \text{Tr}(S_z \mu(t))$, and the total magnetization of the whole isolated system $M_Z = \text{Tr}(M_z \rho(t))$ with $M_z(t) = \sum_{k=1}^{10} S_z^{(k)}$. In all the numerical simulations we use units such that $k_B$ and $\hbar$ are one, and we take $\omega_0$ as the arbitrary energy unit (this implies the time to be expressed in arbitrary unit of $\hbar/\omega_0$).

First le us consider the response to a sudden quench of the external magnetic field of a single pure state. To this aim we have numerically solved the Schrödinger equation for an initial pure state drawn from a RPSE corresponding to an initial typical energy of $E = -2.619$ in the presence of the initial magnetic field $B_i$. In figure (1) we report the evolution of the total magnetization $M_z$: for $\tau < 0$, before the quench, the observable (red line) fluctuates around its equilibrium average (dotted line), $\overline{P}_Z = \text{Tr}(M_z \overline{\rho}) = -1.914$. At $t = t_F$ the external field is instantaneously reduced to $B_f = B_i/5$, the exact evolution of the magnetization (black line) shows an initial relaxation (magnified in the right panel) and then it fluctuates around its new equilibrium average.

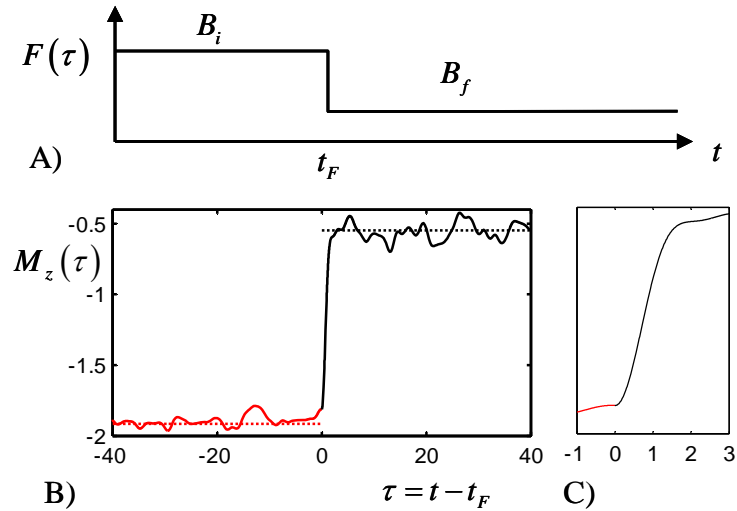

**Figure 1: Response of a quantum pure state describing the state of a system composed of 10 spins $J = 1/2$ to a quench of the external magnetic field. A) Time profile of the applied external field, B) time evolution of the total magnetization, C) magnification of the relaxation after the quench. The parameters used for the simulations shown throughout this paper are the following: we set $\lambda = 0.03$, the particular realization of the pure state has an initial energy of -2.624 and it is drawn from the RPSE with typical energy -2.619 (which corresponds to N=200 populated energy levels).**

The behaviour of the single spin polarization is not shown here but it is qualitative the same of the total magnetization. We emphasize that figure (1) shows the evolution of a single pure state, according to eq.(2), thus no average procedure has been performed. Notice that the statistical theory for the response presented in the previous section does not assume neither imply the existence of a true relaxation dynamics after the quench of the Hamiltonian. Indeed, the evolution showed in figure 1 follows from the exact solution of the unitary dynamics and thus it is completely reversible. In a strictly formal sense there is no irreversible behaviour since both the expectation values of any observable, eq. (3), as well as its statistical average, eq. (48), are quasiperiodic functions for finite dimensional quantum system. However, in the considered case as well as in many other pertinent scenarios [22] recurrences usually occur on time scales which are not physically relevant [23] and thus we shall intend the relaxation dynamics in term of a quasi-irreversible evolution as that showed in figure 1.

In our framework the statistical response is defined by eq. (48) as the average on the initial PSD which describes the equilibrium of the system prior to the change of the external field. In figures 2-3 we show the statistical response (bold line) in terms of the single spin polarization $\bar{m}_z(\tau)$ and the total magnetization $\bar{M}_z(\tau)$, respectively. Together with the statistical response, we also report four responses (coloured lines) of particular pure states characterized by the same set of populations but different phases (notice that this can also be interpreted as the response of a single pure state to the quench occurring at different times). The plots highlight the strong similarity between the statistical response and the response calculated for a single pure state, in particular during the dynamics just after the quench, i.e. during the relaxation toward the new equilibrium. This immediately follows from equation (55) for the subsystem property $m_z(\tau)$, however figure 3 suggests that this similarity is even more stringent for the total magnetization.

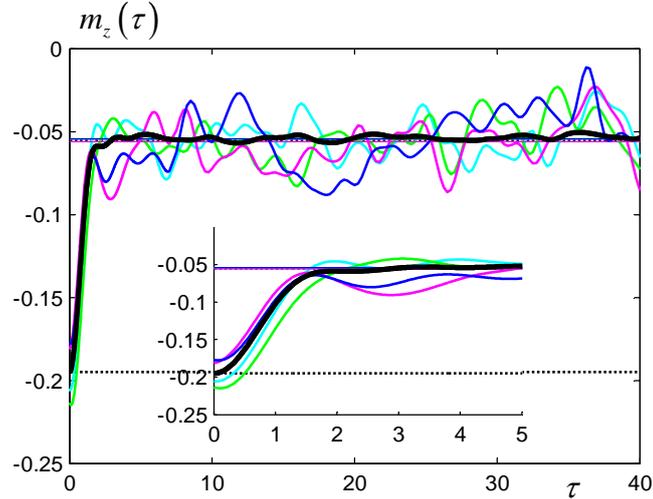

**Figure 2: Time Evolution of the single spin polarization after the quench of the spin Hamiltonian. The coloured lines refer to the response of four different pure states characterized by the same set of populations but different phases, the bold black line corresponds to the statistical response, eq. (48). In the inset the initial relaxation is magnified. The parameters used for the simulation are those reported in Figure 1.**

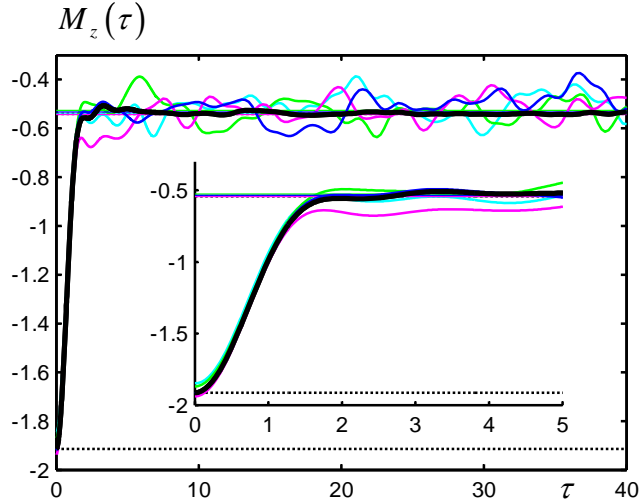

**Figure 3: Time Evolution of the total magnetization after the quench of the spin Hamiltonian. The coloured lines refer to the response of four different pure states characterized by the same set of populations but different phases, the bold black line corresponds to the statistical response, eq. (48). In the inset the initial relaxation is magnified. The parameters used for the simulation are those reported in Figure 1.**

Notice that the calculation of the average response, eq. (48), is operationally equivalent to the calculation of the dynamics of a statistical mixture of pure states, i.e. the time evolution of the statistical density matrix $\bar{\rho} = \sum_n P_n |n\rangle\langle n|$ defined according to eq. (11). The general problem of determining the evolution of statistical density matrices for a system of n spins is of particular practical relevance in the field of magnetic resonance and in particular for NMR experiments connected with the development of ensemble quantum computation [24]. To describe an n-spin ½ system in a mixed state we need a $2^n \times 2^n$ density matrix, but this quickly reaches a storage limit as n increases. Alternatively, the use of wave-functions combined with the methods available for the approximation of the evolution operator [25] overcomes this limitation by using vector of size $2^n$. Within this scheme one first calculates the evolution of each of the wave function $|n\rangle$ which participates to the statistical density matrix, and then the average over all such individual evolutions gives the dynamics of the statistical mixture. Notice that the number of time evolution required for such a calculation scales exponentially with $n$. The idea that the expectation value of a property of interest evaluated during the time evolution of a *single pure states* that are superposition of all the elements of the statistical density matrix can closely mimics the evolution determined by the dynamics of the statistical mixture has been proposed in Refs.[26,27]. It is clear that if an observable shows this type of self-averaging property its evolution can be conveniently determined on the basis of the calculation for a single pure state rather then for $2^n$ pure states. In Ref. [27] it has been proven that this is indeed the case if the initial non equilibrium state has a local character, i.e. an initial perturbing excitation acts only on a small subsystem and we are interest in the evolution of some subsystem properties.

Our calculations show that self-averaging can occur also for collective observables, as the total magnetization shows in figure (3), and in the case of an external force which acts on the whole system, that is, for a global non-equilibrium initial condition. This is indeed a manifestation of the typicality of the responses derived in Section 3. To have an insight into the origin of the self-averaging and thus understanding which parameters can influence it, let us write explicitly the response of a single pure state eq. (46) for the considered quantum quench. Let $\{|f\rangle, f=1....2^n\}$ be the eigen energy basis of the Hamiltonian $H_f$ corresponding to the external field $B_f$, and $\{|n\rangle, n=1....2^n\}$ the eigen energies of the Hamiltonian prior to the quench. The response, for $\tau > 0$, in term of the generic observable $B$ for the initial pure state $|\psi(0)\rangle = \sum_{n=1}^{N} \sqrt{P_n} e^{i\alpha_n} |n\rangle$ is given explicitly by

$$b(\tau) = \text{Tr}(B\rho(t)) = \sum_{ff'} B_{ff'} e^{-i\omega_{ff'} t} \rho_{ff'}(0) \tag{59}$$

whit the initial pure state density matrix elements in the $f$-basis explicitly given by

$$\rho_{ff'}(0) = \langle f|\psi(0)\rangle\langle\psi(0)|f'\rangle = \sum_{nn'} \langle f|n\rangle\langle n'|f'\rangle \sqrt{P_n P_{n'}} e^{i(\alpha_n - \alpha_{n'})}$$
$$= \bar{\rho}_{ff'} + \sum_{n\neq n'} \langle f|n\rangle\langle n'|f'\rangle \sqrt{P_n P_{n'}} e^{i(\alpha_n - \alpha_{n'})} \tag{60}$$

In the last equality $\bar{\rho}_{ff'}$ denotes the elements of the equilibrium density matrix, $\bar{\rho} = \sum_n P_n |n\rangle\langle n|$, in the $f$-basis. By inserting eq. (60) in eq. (59) and by using the definition of statistical response eq. (48) one obtains

$$b(\tau) = \bar{b}(\tau) + \sum_{ff'} B_{ff'} e^{-i\omega_{ff'} \tau} \sum_{nn'\neq n} \langle f|n\rangle\langle n'|f'\rangle \sqrt{P_n P_{n'}} \exp(i(\alpha_n - \alpha_{n'})) \tag{61}$$

The difference between the average dynamics and the dynamics of the single pure state is represented by the second term in eq. (61). Thus, the degree of self-averaging of the observable $B$ depends basically from three factors: i) the degree of randomness of the phases of the coefficients of the initial state $|\psi(0)\rangle$. As already noted in refs [26,27] randomly correlated pure states (i.e. with all the phases distributed homogeneously) are more effective in miming the evolution of the statistical mixture than other initial state of product form. The independence of the phases favours in fact the cancellation of the second term in (61). ii) The Hamiltonian which rules the dynamics, indeed this factor determines the overlaps between the eigenstates in the absence of the perturbation and those in the presence of the external perturbation, i.e. $\langle f|n\rangle$, and of course the frequencies $\omega_{ff'}$. Also the influence of this factor is considered in the analysis of ref 27 which compares two spin systems with different coupling networks (a ladder of spins interacting through an XY Hamiltonian and a "star" spin system in which all the spins interact each other with random distributed intensities). As it is clear from (61) Hamiltonian which are not highly symmetric assure a more efficient self-averaging of the observable because of the interference between the terms evolving with different frequency $\omega_{ff'}$. Finally iii) the self–averaging depends also on the observable through its matrix representation in the final eigenstates. This is indeed the factor which determines the difference in the self averaging properties of the single spin polarization (Figure 2) and the total magnetization (Figure 3). Evidently the matrix structure of the collective observables implies more non-negligible contributions to the summation in eq. (61) with respect to the single spin polarization favouring their mutual cancellation and thus a more efficient self-averaging effect.

We now turn to the dependence of the response on the choice of the initial populations. All the calculations shown in figures 1-3 refer to a particular set of population drawn from a RPSE with an active space of dimension $N = 200$ (corresponding to a typical energy of $E = -2.619$). Do different realizations evolve in a similar manner and toward a similar equilibrium? In figure 4-5 we report the statistical responses of the local and the total magnetization, respectively, for different sets of populations (coloured lines), together with its RPSE average (bold line). The equilibrium averages of such observables prior to the quench are also reported (dotted lines). They are realizations from the corresponding RPSE distribution which is showed in the right-bottom panel. The typicality of the response function with respect to different choices of the initial population in the ensemble are well evident, again typicality holds for the subsystem observable and also for the total magnetization.

The last issue we shall consider is the nature of the new equilibrium reached after the quench of the magnetic field. In particular, the initial states are drawn from a RPSE with a typical energy of $E = -2.619$ in the presence of the initial magnetic field, after the quench the typical energy becomes $E = -1.080$. However, because of the change of the eigenenergy-basis, the populations after the quench are not distributed according to the RPSE statistics. Since such new population sets determine the equilibrium average value of the observables after the quench, we wonder if such a value agrees with the typical value of the same observable in a RPSE characterized by the same typical energy of $E = -1.080$. In order to investigate this point we have considered a RPSE in an active space of dimension $N = 384$ which gives the required typical energy. The distributions of the observables in such an ensemble are shown in the right upper panel of figures 4-5. The RPSE typical values $\langle \bar{m}_z \rangle = -0.054$ and $\langle \bar{M}_z \rangle = -0.533$ are in surprisingly good agreement with the values obtained from the dynamics. This is a particularly important point as long as we know that the RPSE statistics assure a description of the equilibrium which is consistent with thermodynamics [11]. Thus, for the model at hand we can conclude that not only the considered observables tend to equilibrate at a given value, but such a value is also thermodynamically consistent with the total energy of the system. In other words we can say that the system *thermalize* [6].

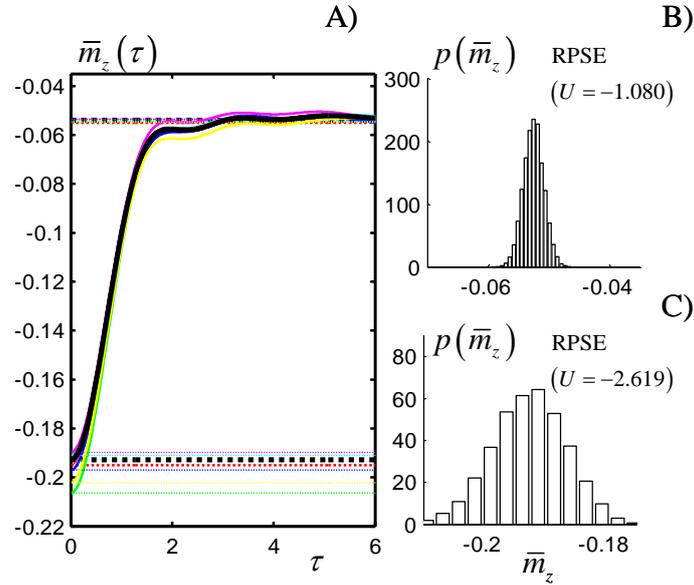

**Figure 4: A) Statistical response in term of single spin polarization after the quench of the Hamiltonian for different realizations of the RPSE (coloured lines) and the corresponding RPSE average response (black bold line). The dotted lines correspond to the equilibrium value of the single spin polarization, before and after the quench, for each particular realization of the ensemble. The parameters used for the simulation are those reported in Figure 1. On the right panels the RPSE distribution of the same observable is depicted for C) the initial RPSE and B) the RPSE which corresponds to the typical energy after the quench.**

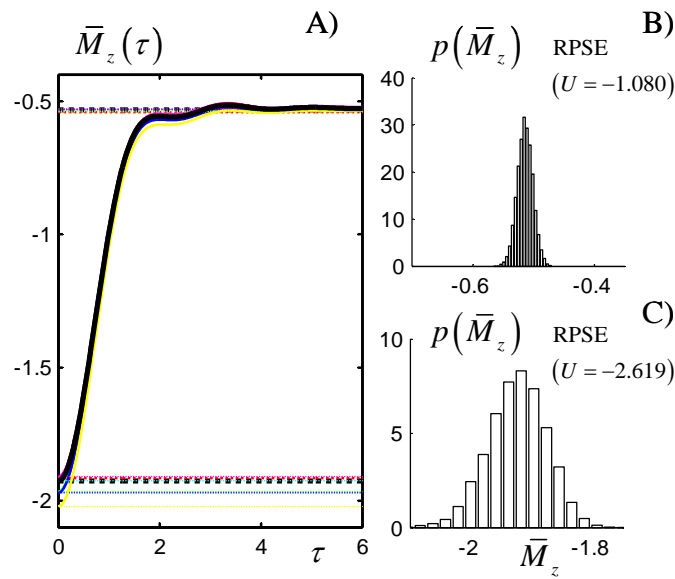

**Figure 5: A) Statistical response in term of total magnetization after the quench of the Hamiltonian for different realizations of the RPSE (coloured lines) and the corresponding RPSE average response**

(black bold line). The dotted lines correspond to the equilibrium value of the total magnetization, before and after the quench, for each particular realization of the ensemble. The parameters used for the simulation are those reported in Figure 1. On the right panels the RPSE distribution of the same observable is depicted for C) the initial RPSE and B) the RPSE which corresponds to the typical energy after the quench.

**IV. Linear Response**

The response theory in the linear regime is a cornerstone of non equilibrium statistical mechanics. Thus we shall derive it as a particular case of the general treatment given in the previous section and comment about its relation with the standard Kubo's formulation [4]. Despite the successes of linear response theory, its microscopic foundation has been subject of vivacious debates [28,29]. Since the salient feature of the framework presented in the previous section is a clear distinction between the mechanical level and the statistical assumptions, our derivation of the linear response shed light on some problematic issues of the standard formulation [30,31].

We need to derive for the response $B(t_F, \tau)$ eq. (46) the linear contribution with respect the external force $F(\tau)$. Let us consider the order expansion of the unitary evolution operator $\hat{U}(\tau)$ with respect to the force:

$$\hat{U}(\tau) = \hat{U}^{(0)}(\tau) + \hat{U}^{(1)}(\tau) + \hat{U}^{(2)}(\tau) + \cdots \tag{62}$$

where $\hat{U}^{(n)}(\tau)$ is of the order $F(\cdot)^n$. From the evolution equation for $\hat{U}(\tau)$

$$i\hbar \frac{\partial}{\partial \tau} \hat{U}(\tau) = \left[\hat{H} + F(\tau)\hat{A}\right]\hat{U}(\tau) \tag{63}$$

one derives an hierarchy of equations of increasing order, whose explicit solutions for the first two orders are

$$\hat{U}^{(0)}(\tau) = \exp\left\{-i\hat{H}\tau/\hbar\right\} \qquad \hat{U}^{(1)}(\tau) = -(i/\hbar)\hat{U}^{(0)}(\tau) \int_0^\tau d\tau' F(\tau')\hat{A}^{(0)}(\tau') \tag{64}$$

where

$$\hat{A}^{(0)}(\tau) := \hat{U}^{(0)}(\tau)^+ \hat{A} \hat{U}^{(0)}(\tau) \tag{65}$$

is the Heisenberg form of operator $\hat{A}$ calculated with the evolution operator of zero-th order. By substitution of eq. (62) into eq. (47), one can calculate the order expansion of $\hat{B}(\tau)$

$$\hat{B}(\tau) = \hat{B}^{(0)}(\tau) + \hat{B}^{(1)}(\tau) + \hat{B}^{(2)}(\tau) + \cdots \tag{66}$$

where $\hat{B}^{(o)}(\tau) = \hat{U}^{(o)}(\tau)^\dagger \hat{B} \hat{U}^{(o)}(\tau)$ would describe the evolution of $\hat{B}(\tau)$ in the absence of the force, and with the following explicit form for the first order correction

$$\hat{B}^{(1)}(\tau) = -(i/\hbar) \int_0^\tau d\tau' F(\tau') \left[\hat{B}^{(0)}(\tau), \hat{A}^{(0)}(\tau')\right] \tag{67}$$

Finally the linear response, that is the first order contribution in eq. (46), is derived as

$$\delta B(t_F,\tau) := B(t_F,\tau) - B(t_F,\tau)\big|_{F(\cdot)=0} =$$
$$= -(i/\hbar)\int_0^\tau d\tau' F(\tau')\text{Tr}\left\{\left[\hat{B}^{(0)}(\tau),\hat{A}^{(0)}(\tau')\right]\hat{\rho}(t_F)\right\} \quad (68)$$

Notice that the linear response still depends through $\hat{\rho}(t_F)$ on the particular state of the system prior the application of the force. Such a dependence is eliminated if the average on the fluctuations is performed

$$\overline{\delta B(\tau)} = \int_0^\tau d\tau' F(\tau')\varphi_{\hat{B},\hat{A}}(\tau-\tau') \quad (69)$$

with the following response function

$$\varphi_{\hat{B},\hat{A}}(t) = -(i/\hbar)\text{Tr}\left\{\left[\hat{B}^{(0)}(t),\hat{A}\right]\overline{\hat{\rho}}\right\} = -(i/\hbar)\text{Tr}\left\{\left[e^{i\hat{H}t/\hbar}\hat{B}e^{-i\hat{H}t/\hbar},\hat{A}\right]\overline{\hat{\rho}}\right\} \quad (70)$$

Notice that in the case we are looking at the response of a small subsystem, then the vanishing amplitude of fluctuations assures that the difference between the response of a single system eq. (68) and the average response given by eq. (69) at a given time $\tau$ tends to zero as the size of the global system increases. This result can be interpreted on the light of the long debate about the microscopic origin of linear response theory triggered by the well known Van Kampen observation [28] pointing out the difference between the linearity of microscopic motion and the macroscopic linearity arising from some randomization procedure [30,32,33]. The vanishing of fluctuations for a subsystem supports the validity of the linear response even for individual trajectory, without invoking the constitutive role of a randomization procedure [30,34].

Kubo's formulation of linear response corresponds to assuming the canonical form of the equilibrium density matrix. Although such a particular choice leads to the celebrated fluctuation-dissipation relations [35] it is rather inconsistent with the equation of motion for an isolated system which are the starting point of the entire Kubo's derivation and that does not account for any dissipation. This was the basic concern of Van Vliet which, for example, developed the idea that linear response theory should rely either on a master equation approach or on some projective procedure [36,37]. More recently, linear response has been analysed in the framework of the standard microcanonical statistics [38] by showing that fluctuation-dissipation relations are not confined to the canonical set up. In our formulation, eq. (70), the equilibrium density matrix corresponds to the specific PSD considered, eq. (11), and in this sense it is more general then the analogue forms based on some specific statistical ensemble. Moreover it should be stressed that the response function for a subsystem will assume a typical value in correspondence of its RPSE average

$$\left\langle\varphi_{\hat{B},\hat{A}}(t)\right\rangle = -(i/\hbar)\text{Tr}\left\{\left[e^{i\hat{H}t/\hbar}\hat{B}e^{-i\hat{H}t/\hbar},\hat{A}\right]\left\langle\overline{\hat{\rho}}\right\rangle\right\} \quad (71)$$

as a consequence of the analysis of the previous section.

Notice that within the RPSE statistics (and also within the standard microcanonical set up), the response is determined exclusively by the transition frequency between the states populated at equilibrium $E_n \leq E_{max}$ and those which are not populated at equilibrium $E_n > E_{max}$. To make this evident let us assume $B = A$, then eq. (71) can be written explicitly as

$$\langle \phi_{BA}(t) \rangle = 2 \sum_{\substack{n \leq n_{max} \\ n' > n_{max}}} |A_{nn'}|^2 \sin(\omega_{nn'} t) \tag{72}$$

Such a feature is not recovered in the standard canonical formulation of the linear response function.

An interesting point to analyze is under which condition the typical response of eq. (71) reduces to the Kubo's canonical formulation. Let us assume that both the observable and the coupling with the external force are described by operators of a subsystem, $\hat{B} = \hat{b} \otimes \hat{1}_E$, $\hat{A} = \hat{a} \otimes \hat{1}_E$. Furthermore we consider a subsystem large enough such that the Hamiltonian coupling the subsystem with the environment can be neglected in the evolution of $\hat{B}^{(0)}(t)$. Then the RPSE averaged response function eq. (71) can be specified as

$$\langle \varphi_{\hat{B},\hat{A}}(t) \rangle = -(i/\hbar) \text{Tr}_S \left\{ \left[ e^{i\hat{H}_S t/\hbar} \hat{b} e^{-i\hat{H}_S t/\hbar}, \hat{a} \right] \langle \overline{\hat{\mu}} \rangle \right\} \tag{73}$$

where $\hat{H}_S$ is the subsystem Hamiltonian and $\langle \hat{\mu} \rangle$ is the averaged reduced density matrix for the subsystem that, in the large size limit of the overall system, has the canonical form eq. (23). This is precisely the Kubo form for the response function.

**Conclusion**

In this paper we have presented a statistical theory of the response of quantum pure states. The main goal is the definition of a statistical framework in which the probability distribution used to compute the pertinent averages has a precise physical meaning and does not require the introduction of *ad hoc* initial density matrices describing the quantum system as a very peculiar statistical mixture of quantum states. A second main result is the proof of the existence of a typical response, in the meaning that the overwhelming majority of quantum pure states belonging to a given RPSE respond in almost the same way to an external perturbation.

The typical response to a quench of the magnetic field has been examined for a system composed of n=10 spins ½. Some remarks on the results obtained for this specific model system are in order: we have shown that after the quench both the total magnetization and the single spin polarization tend to relax toward a new equilibrium value. Notice that we will not couple this model to any other degrees of freedom (any external "bath"), thus this isolated interacting quantum system itself can serve as its own heat bath in order to relax to equilibrium after the perturbation of an external field. For the parameters used in the calculations presented in this paper we have also shown that the equilibrium reached after the quench of the Hamiltonian is compatible with the description of a *thermal* equilibrium.

Finally, linear response theory has been derived from our treatment of the response to generic external fields. The standard form of the response function has been obtained by specifying the physical meaning of the underlying statistical assumptions. Moreover we have discussed our result in connection with the canonical formulation of Kubo [4]. Although beyond the scope of the present work a general analysis of the relation between the canonical, the microcanonical and our more general linear response theory would be of great interest and value.

**References and Note**